\documentclass{appolb}
\usepackage{graphicx}
\begin{document}
\title{Microscopic Calculations of Stellar Weak Rates for sd- and fp-Shell Nuclei for Astrophysical Applications.\thanks{Presented at XXXIII Zakopane School Of Physics 1998.}}
\author{Jameel-Un-Nabi and H. V. Klapdor-Kleingrothaus
\address{Max-Planck-Institut f\"ur Kernphysik, Postfach 103980, 69029 Heidelberg}}
\maketitle
\begin{abstract}
Proton-netron quasiparticle RPA is used for the first time to calculate weak interaction rates for $\mathit{sd}$- and $\mathit{fp}$-shell nuclei at high temperatures and densities. The calculated rates take into consideration the latest experimental energy levels and ft value compilations. Particle emission processes from excited states are taken into account. The calculation is done for 700 nuclei with mass number ranging from 18 to 100. The astrophysical applications of the calculated rates are highlighted.
\end{abstract}
  
\section{Introduction}
In the course of development of a star, the weak interaction has several crucial effects. They initiate the gravitational collapse of the core of a massive star triggering a supernova explosion, play a key role in neutronisation of the core material, and, affect the formation of heavy elements above iron via the r-process at the final stage of the supernova explosion. The weak interaction also largely determines the mass of the core, and thus the strength and fate of the shock wave formed by the supernova explosion.

Precise knowledge of the $\beta$ decay of neutron-rich nuclei is crucial to an understanding of the r-process. Both the element distribution on the r-path, and the resulting final distribution of stable elements are highly sensitive to the $\beta$ decay properties of the neutron-rich nuclei involved in the process. This was first pointed out by Klapdor [1,2,3] (see also [4]).There are about 6000 nuclei between the $\beta$ stability line and the neutron drip line. Most of these nuclei cannot be produced in terrestrial laboratories and one has to rely on theoretical extrapolations in respect of beta decay properties. Calculations of beta decay rates for all nuclei far from stability by microscopic nuclear theory were first performed by Klapdor et al. [5], and then complemented and refined by Staudt et al. [6,7] and Hirsch et al. [8]. Recent studies by Homma et al. [9] and  M\"oller et al. [10] have shown that the best extrapolations to nuclei far from stability to date still are given by [6]. All these weak interaction rates led to a better understanding of the r-process. However there was a need to go to domains of high temperature and density scales where the weak interaction rates are of decisive importance in a study of the stellar evolution.

There are many interesting problems in astrophysics which require $e^{\pm}$ -capture and $\beta^{\pm}$ -decay rates as input parameters. These rates are used in numerical simulations for nuclear astrophysical problems. Some processes where $\beta$ transitions occur at high temperatures and densities (T$ \sim$ 10$^{9}$ K, $\rho Y_{e}$ $\geq$ 10$^{3}$ gcm$^{-3}$) include the initiation of the collapse of the O + Ne + Mg core of 8-10 M$_{\odot}$ stars [11], the growth of the mass of the iron core of M $\geq$ 10 M$_{\odot}$ stars -- the element synthesis problem, like the synthesis of iron-group elements in carbon-detonation supernova models [12], the hot CNO-cycle [13,14], where at temperatures beyond 10$^{8}$ K and densities beyond 1 gcm$^{-3}$, $\beta^{+}$ decays are much slower than thermonuclear reactions. Some other examples include the s-process [15,16], and the p-process [17,18].

Proton-neutron quasiparticle random-phase-approximation (pn-QRPA) has been shown to be a good microscopic theory for the  calculation of beta decay half-lives. Bender et al. [19] and Staudt et al. [6,7] used the QRPA to calculate the $\beta$$^{-}$ decay half-lives of nuclei far from stability and obtained good agreement with experimental decay rates. Muto et al. [20] then extended this model to treat transitions from nuclear excited states. Keeping in view the success of pn-QRPA theory in calculating terrestrial decay rates, we used this extended model to calculate for the first time the weak interaction rates in stellar matter using pn-QRPA theory. The main advantage of using this theory is that it enables us to calculate also the weak interaction rates of $\mathit{fp}$-shell nuclei. This work is the first calculation of weak rates for $\mathit{fp}$-shell nuclei in hot and dense matter taking into account details of the nuclear structure and presenting the rates over a wide range of temperature and density grid points. We took into consideration for the first time the effect of beta-delayed particle emission in this type of calculations. The calculated weak interaction rates for about 700 nuclei (A = 18 to 100) can be obtained as files on a magnetic tape from the authors on request.

In sections 2 of this paper we present briefly the formalism used. For details of formalism and calculations we refer to our paper [21]. Section 3 discusses some astrophysical applications of the calculated rates. Section 4 summerizes this work.
\newpage
\section{Formalism}
The weak decay rate from the $\mathit{i}$th state of the parent to the $\mathit{j}$th state of the daughter nucleus is given by \footnote{Throughout section 2 we use natural units $(\hbar=c=m_{e}=1)$, unless otherwise stated, where $m_{e}$ is the electron mass.}
\begin{equation}
\lambda_{ij} =ln2 \frac{f_{ij}(T,\rho,E_{f})}{(ft)_{ij}}
\end{equation}
where $t_{ij}$ is the half-life and $f$ is the Fermi integral, both being related to the reduced transition probability $B_{ij}$ of the nuclear transition by
\begin{equation}
(ft)_{ij}=D/B_{ij}
\end{equation}
The $D$ appearing in Eq. (2) is a compound expression of physical constants,
\begin{equation}
D=\frac{2ln2\hbar^{7}\pi^{3}}{g_{V}^{2}m_{e}^{5}c^{4}}
\end{equation}
and,
\begin{equation}
B_{ij}=B(F)_{ij}+(g_{A}/g_{V})^2 B(GT)_{ij}
\end{equation}
where B(F) and B(GT) are reduced transition probabilities of the Fermi and Gamow-Teller (GT) transitions respectively,
\begin{equation}
B(F)_{ij} = \frac{1}{2J_{i}+1} \mid<j \parallel \sum_{k}t_{\pm}^{k} \parallel i> \mid ^{2}
\end{equation} 
\begin{equation}
B(GT)_{ij} = \frac{1}{2J_{i}+1} \mid <j \parallel \sum_{k}t_{\pm}^{k}\vec{\sigma}^{k} \parallel i> \mid ^{2}
\end{equation}

In Eq. (6), $\vec{\sigma}^{k}$ is the spin operator and $t_{\pm}^{k}$ stands for the isospin raising and lowering operator. We take the value of D=6295 s and the ratio of the axial-vector $(g_{A})$ to the vector $(g_{V})$ coupling constant as 1.254. The calculation of nuclear matrix elements is dealt with in section 3.

The phase space integral $(f_{ij})$ is an integral over total energy,
\begin{equation}
f_{ij} = \int_{1}^{w_{m}} w \sqrt{(w^{2}-1)} (w_{m}-w)^{2} F(\pm Z,w) (1-G_{\mp}) dw
\end{equation}
for electron (\textit{upper signs}) or positron (\textit{lower signs}) emission, or by
\begin{equation}
f_{ij} = \int_{w_{l}}^{\infty} w \sqrt{(w^{2}-1)} (w_{m}+w)^{2} F(\pm Z,w) G_{\mp} dw 
\end{equation}
for continuum positron (\textit{lower signs}) or electron (\textit{upper signs}) capture.

In Eqs. (7) and (8), $w$ is the total kinetic energy of the electron including its rest mass, $w_{l}$ is the total capture threshold energy (rest+kinetic) for positron (or electron) capture. One should note that if the corresponding electron (or positron) emission total energy, $w_{m}$, is greater than -1, then \\$w_{l}=1$, and if it is less than or equal to 1, then $w_{l}=\mid w_{m} \mid$. $w_{m}$ is the total $\beta$- decay energy,
\begin{equation}
w_{m} = m_{p}-m_{d}+E_{i}-E_{j}
\end{equation}
where $m_{p}$ and $E_{i}$ are mass and excitation energies of the parent nucleus, and $m_{d}$ and $E_{j}$ of the daughter nucleus, respectively.

$G_{+}$ and $G_{-}$ are the positron and electron distribution functions, respectively. Assuming that the electrons are not in a bound state, these are the Fermi-Dirac distribution functions,
\begin{equation}
G_{-} = [exp (\frac{E-E_{f}}{kT})+1]^{-1}
\end{equation}
\begin{equation}
G_{+} = [exp (\frac{E+2+E_{f}}{kT})+1]^{-1}
\end{equation}
Here $E=(w-1)$ is the kinetic energy of the electrons, $E_{f}$ is the Fermi energy of the electrons, $T$ is the temperature, and $k$ is the Boltzmann constant.

In our calculations, the inhibition of the final neutrino phase space is never appreciable enough that neutrino (or anti-neutrino) distribution functions had to be taken into consideration. $F(\pm Z,w)$ are the Fermi functions and are calculated according to the procedure adopted by Gove and Martin [22].

The number density of electrons associated with protons and nuclei is $\rho Y_{e} N_{A}$, where $\rho$ is the baryon density, $Y_{e}$ is the ratio of electron number to the baryon number, and $N_{A}$ is the Avogadro's number.
\begin{equation}
\rho Y_{e} = \frac{1}{\pi^{2}N_{A}}(\frac {m_{e}c}{\hbar})^{3} \int_{0}^{\infty} (G_{-}-G_{+}) p^{2}dp 
\end{equation}
where $p=(w^{2}-1)^{1/2}$ is the electron or positron momentum, and Eq. (12) has the units of \textit{moles $cm^{-3}$}. This equation is used for an iterative calculation of Fermi energies for selected values of $\rho Y_{e}$ and $T$.

The rate per unit time per nucleus for any weak process is then given by
\begin{equation}
\lambda = \sum_{ij}P_{i} \lambda_{ij}
\end{equation}
Here $P_{i}$ is the occupation probability of state i, calculated on the assumption of thermal equilibrium. We carry out this summation over all initial and final states until satisfactory convergence in the rate calculations is achieved.

The neutrino energy loss rates are calculated using the same formalism except that the phase space integral is replaced by
\begin{equation}
f_{ij}^{\nu} = \int_{1}^{w_{m}} w \sqrt{(w^{2}-1)} (w_{m}-w)^{3} F(\pm Z,w) (1-G_{\mp}) dw
\end{equation}
for electron (\textit{upper signs}) or positron (\textit{lower signs}) emission, or by
\begin{equation}
f_{ij}^{\nu} = \int_{w_{l}}^{\infty} w \sqrt{(w^{2}-1)} (w_{m}+w)^{3} F(\pm Z,w) G_{\mp} dw  
\end{equation}
for continuum positron (\textit{lower signs}) or electron (\textit{upper signs}) capture.

We calculate the proton (neutron) energy rate from the daughter nucleus, whenever $S_{p(n)}$ $<$ $S_{n(p)}$, by
\begin{equation}
\lambda^{p(n)} = \sum_{ij}P_{i}\lambda_{ij}(E_{j}-S_{p(n)})
\end{equation}
for all $E_{j} > S_{p(n)}$, whereas for all $E_{j} \leq S_{p(n)}$ we calculate the $\gamma$ heating rate,
\begin{equation}
\lambda^{\gamma} = \sum_{ij}P_{i}\lambda_{ij}E_{j}
\end{equation}

For the details of the formalism and calculations we refer to [21].
\section{Astrophysical applications of calculated rates}
\subsection{Evolution of 8-10 M$_{\odot}$ stars}
Several circumstantial arguments point to the formation of the third r-process peak at A $\sim$ 190 in 8-10 M$_{\odot}$ stars [23, and the references therein]. These low-mass supernovae might produce an r-process in a prompt explosion and a subsequent r-process in a neutrino driven wind. In [23] it is strongly argued that a prompt explosion of an O+Mg+Ne core may produce a reasonable r-process peak at A $\simeq$ 190, as well as r-process elements with Z $\geq$ 56. This prompt explosion was also calculated for the collapse of a O+Ne+Mg core by Hillebrandt et al. [24]. However other authors like Mayle and Wilson [25] found a delayed explosion due to late time neutrino heating for these cores. 

Since these low-mass supernova stars are very interesting from astrophysical viewpoint, it is desirable to know the fate of these cores much more reliably. In their final stages of evolution the 8-10 M$_{\odot}$ stars undergo electron captures on $^{24}$Mg and $^{20}$Ne. Core collapse begins as the pressure of degenerate electrons starts decreasing. At the same time these reactions liberate entropy which leads to an increase in the core temperature and oxygen deflagration is induced. Therefore, the star will either collapse to leave a neutron star behind or be totally disrupted, depending on which effect is dominant. The critical quantity to determine which process is dominant is the density at which oxygen ignition, $\rho_{ign}$, starts; if $\rho_{ign}$ is low (high), oxygen deflagration (collapse of the core) dominates the other process. There are two main uncertainties involved in order to determine  $\rho_{ign}$; one is the treatment of the semiconvective layer and the other is the uncertainty in the electron capture rates.

Miyaji and Nomoto [26] investigated $\rho_{ign}$ using the electron capture rates from the gross theory and reported in their paper that  it is very unlikely that oxygen burning is ignited by electron capture on $^{24}$Mg and $^{24}$Na ($\rho \sim 4 \times 10^{9}$ gcm$^{-3}$). However they reported that it is likely that at $\rho \sim  10^{10}$ gcm$^{-3}$ electron capture on $^{20}$Ne ignites the oxygen burning and $\rho_{ign}$ is not low enough for complete disruption of stars. 

The temperature and density of the O+Ne+Mg core at this relevant stage are:\\
logT = 8.4 - 9.4 (in Kelvins)\\
log$\rho$ = 9.3 - 10.4 (in gcm$^{-3}$).\\
For the A = 24 systems, if convection is neglected, the electron capture process stops when all $^{24}$Na is exhausted. This occurs because there is no supply of fresh material in the central region. However, if convection is considered, fresh $^{24}$Mg (hence $^{24}$Na) is always supplied in the convective core. As a result the entropy production continues to the higher density where gross theory electron capture rates differ from those of Oda et al. [27] by an order of magnitude which in turn differ from our calculated electron capture rates by an order of magnitude. In other words, for the relevant temperature and density scales, our electron capture rates of $^{24}$Mg differ by two orders of magnitude in comparison with the electron capture rates of the gross theory (see figure ~1).
\begin{figure}[h!]
	\centering
	\begin{minipage}{0.32\textwidth}
		\centering
		\includegraphics[width=6.0cm]{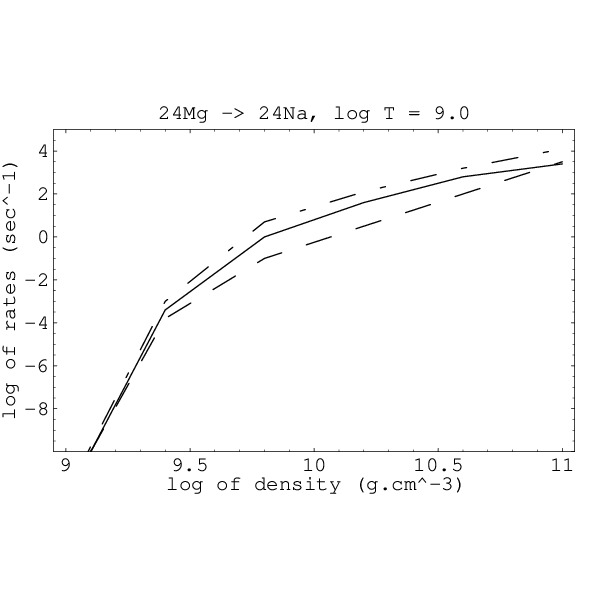}
	\end{minipage}
	\begin{minipage}{0.32\textwidth}
		\centering
		\includegraphics[width=6.0cm]{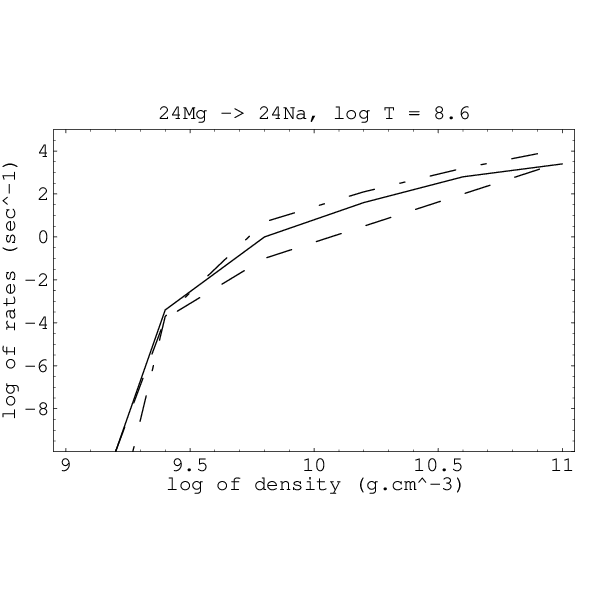}
	\end{minipage}
	\begin{minipage}{0.32\textwidth}
		\centering
		\includegraphics[width=6.0cm]{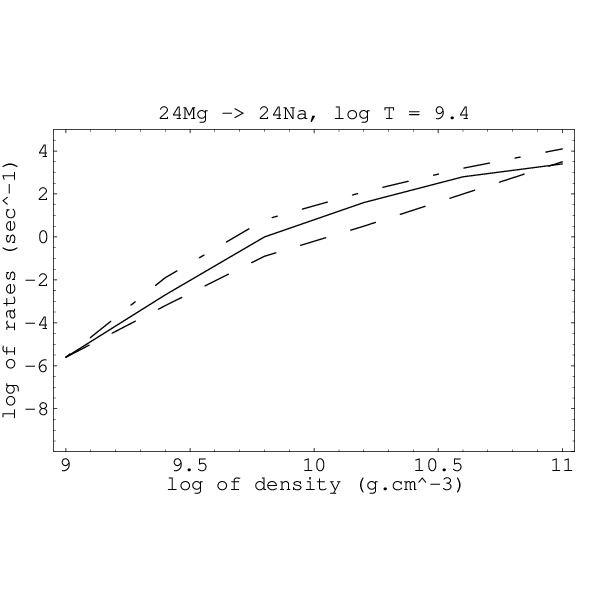}
	\end{minipage}
	\caption{\footnotesize Electron capture rates for $^{\mathrm{24}}$Mg as a function of $\rho Y_{e}$, at log T = 8.6, 9.0, and 9.4. The dot-dashed line represents our rates, the dashed line depicts the rates of Fuller et al. [28], while the solid line shows the results of Oda et al. [27].}
\end{figure}

Since the heating rate also increases in this density region, there is a possibility that the oxygen ignition begins before the electron capture on $^{\mathrm{24}}$Ne starts. In order to obtain the quantitative amount of change of $\rho_{ign}$ using our calculated capture rates, a precise calculation of the evolution of the O+Ne+Mg core is necessary.
\subsection{Search for important weak interaction nuclei in stellar evolution of massive stars.}
The rate of change of $Y_{e}$ is important for stellar evolution. The nuclei which cause the largest changes in $Y_{e}$ are neither the most abundant ones nor the ones with the strongest rates. Knowledge of both abundances \textit{and} rates of nuclei are required for determining the most important electron capture or beta-decay nuclei for given values of $\rho$, $T$, and $Y_{e}$. The rate of change of $Y_{e}$ is defined as [29]:
\begin{equation}
\dot{Y_{e}}^{ec(bd)} (k) = -(+) \frac{X_{k}}{A_{k}} \lambda_{k}^{ec(bd)},
\end{equation}
where the abundance of the kth nucleus is given by:
\begin{equation}
X_{k}=\frac{G(Z,A,T)}{2}(\frac{\rho N_{A}}{2} \lambda^{3})^{A-1}A^{5/2}X_{n}^{N}X_{p}^{P}\\exp[\frac{Q_{k}}{k_{B}T}].
\end{equation}
Here $ G(Z,A,T)$ is the nuclear partition function, $N_{A}$ is the Avagadro number, $k_{B}$ is the Boltzmann constant, $Q_{k}$ is the binding energy, and $\lambda = (\frac{h^{2}}{2\pi m_{H}k_{B}T})^{1/2}$ is the thermal wavelength. $X_{n}$ and $X_{p}$ are determined using the mass and charge conservation of the distribution of nuclei. Hence $X_{k}$, for a nucleus k, can be calculated. Substituting  $\lambda_{k}^{ec(bd)}$ by our calculated rates will allow us to calculate Eq. (18) for all nuclei of interest. This calculation may then be compared with earlier works [29, 30, 31, 32] to check if the listings of the important nuclei changes. Such ranked lists of nuclei may be helpful to prioritize more detailed studies on individual nuclei.
\section{Summary}
Weak interaction rates of $\mathit{sd}$-shell nuclei in stellar matter are calculated for the first time using pn-QRPA theory. This theory which proved to be a successful theory for calculation of terrestrial rates promises to be a good candidate for stellar rates. Some of the astrophysical applications of the calculated rates are discussed. It is shown that evolution of the cores of 8-10 M$_{\odot}$ stars and the associated r-process deserve much greater quantitative study.

\end{document}